\documentclass[pre,superscriptaddress]{revtex4}

\newcommand{\abs}[1]{\mathopen|#1\mathclose|}
\usepackage{epsfig}
\usepackage{caption}
\usepackage{placeins}
\usepackage{soul}
\usepackage{amsfonts}
\usepackage{amssymb}
\usepackage{amsmath}
\usepackage{setspace}
\usepackage{version}
\usepackage{graphicx}
\usepackage{color}

\begin{document}
\title{Elasticity and mechanical instability of charged lipid bilayers
in ionic solutions}

\author{Yotam Y.~Avital}

\affiliation{Ilse Katz Institute for Nanoscale Science and Technology,
Ben Gurion University of the Negev, Be'er Sheva, 84105 Israel}

\affiliation{Department of Biomedical Engineering, Ben Gurion
University of the Negev, Be'er Sheva, 84105 Israel}

\author{Niels Gr{\o}nbech-Jensen}

\affiliation{Department of Mechanical and Aerospace Engineering,
University of California, Davis, CA 95616}

\affiliation{Department of Mathematics, University of California,
Davis, CA 95616}

\author{Oded Farago}

\affiliation{Ilse Katz Institute for Nanoscale Science and Technology,
Ben Gurion University of the Negev, Be'er Sheva, 84105 Israel}

\affiliation{Department of Biomedical Engineering, Ben Gurion
University of the Negev, Be'er Sheva, 84105 Israel}

\begin{abstract} 
  We use coarse-grained Monte Carlo simulations to study the elastic
  properties of charged membranes in solutions of monovalent and
  pentavalent counterions. The simulation results of the two cases
  reveal trends opposite to each other. The bending rigidity and
  projected area increase with the membrane charge density for
  monovalent counterions, while they decrease for the pentavalent
  ions. These observations can be related to the counterion screening
  of the lipid charges. While the monovalent counterions only weakly
  screen the Coulomb interactions, which implies a repulsive Coulomb
  system, the multivalent counterions condense on the membrane and,
  through spatial charge correlations, make the effective interactions
  due the charged lipids attractive. The differences in the
  elastic properties of the charged membranes in monovalent and
  multivalent counterion solutions are reflected in the mechanisms
  leading to their mechanical instability at high charge densities. In
  the former case, the membranes develop pores to relieve the
  electrostatic tensile stresses, while in the latter case, the
  membrane exhibit large wavelength bending instability.
\end{abstract}
\maketitle

Lipid bilayers are fundamental to all living cells and participate
in diverse biological processes \cite{Alberts2002}. Their elasticity is
traditionally treated withing the framework of the Helfrich effective
Hamiltonian relating the elastic energy density (per unit area $dS$)
of a thin sheet to the local principle curvatures $c_1$ and $c_2$
\cite{helf1973}:
\begin{equation}
  {\cal H} = \int_{S} \left [ \gamma + \frac{1}{2} \kappa \left ( c_1 +
  c_2 -2c_0\right)^2 + \kappa_G c_1 c_2 \right ] dS,
  \label{eq:helf0}
\end{equation}
where $\gamma$, $\kappa$, and $\kappa_G$ denote the surface tension,
bending rigidity, and saddle-splay modulus, respectively, and $c_0$ is
the spontaneous curvature. The elastic moduli appearing in
Eq.~(\ref{eq:helf0}) are primarily governed by the short-range
intermolecular forces between the lipids \cite{Pincus2004}. Biological
membranes, however, are often charged and suspended in ionic
solutions, and the long-range electrostatic interactions are expected
to contribute to $\kappa$ and $\kappa_G$.  Solutions of the
Poisson-Boltzmann (PB) equation for spherical, cylindrical, and
sinusoidal geometries suggest that $\kappa$ increases with the surface
charge density \cite{Mean1995}. This trend of increase in $\kappa$,
predicted by the mean field theory, holds true for both weak and
strong electrolytes, irrespective of the degree of electric coupling
due the two monolayers. Going beyond mean field, by considering charge
density fluctuations and their spatial correlations, has led to
several opposing predictions that $\kappa$ may be reduced in the
presence of charges \cite{Pincus1998, Shklovskii1999,
Netz2001,Sung2002}. Spatial charge correlations effects (which are
missing in the mean field description) become significant when bilayer
membranes are suspended in solutions with multivalent counterions
\cite{Safran2002,Netz2002}. The phenomenon most associated with these
effects is the attraction between similarly charged objects
\cite{Bloomfield1996,Gelbert1997,Beardmore1998}, which is found in
many biological systems, e.g., in cell-cell adhesion and DNA
condensation. The reduction in $\kappa$ of bilayer membranes is a less
commonly recognized phenomenon involving both charge-charge and
\emph{height-charge}\/ correlations. The details of this effect are
not fully understood, but it is generally believed to be small and,
thus, hard to measure experimentally. Also, the attractive
electrostatic interactions between multivalent counterions, which
presumably lead to a decrease in $\kappa$, may be overshadowed by
other intermolecular interactions, especially when screened by added
salt. It has, in fact, been suggested \cite{Pincus1998,Sung2002} that
when the Debye screening length falls below the typical spacing
between the multivalent counterions, the change in $\kappa$ becomes
positive rather than negative (as predicted by mean field theory).
Here, we approach this topic through computer simulations of
fluctuating bilayers, which allow us to explore the effect in the
absence of added salt and other influences. Our results support the
picture that spatial correlation in the charge density due to the
presence of highly multivalent counterions tend to soften membranes
and reduce their bending rigidity. To the best of knowledge, this is
the first simulation study to report this phenomenon, with the
exception of the work of Fleck and Netz \cite{fleck} who investigated
the enhancement of short wavelength protrusion modes (to be
distinguished from the long wavelength bending modes discussed here)
due to electrostatic repulsion between the membrane charges.

The bending modulus of symmetric bilayers [$c_0=0$ in
Eq.~(\ref{eq:helf0})] can be derived by considering the Fourier space
representation of the Helfrich effective Hamiltonian. For a weakly
fluctuating membrane, the surface $S$ can be described within the
Monge gauge by the height function $h(\vec{r})$ above a square flat
reference surface of linear size $L$. Upon introducing the Fourier
transform of $h(\vec{r}): h_{\vec q} = (l/L)\sum_{i=1}^{(L/l)^2} h
\left ( \vec r_i \right ) e^{i \vec q \cdot \vec r}$, where the sum
runs over a set of $(L/l)^2$ grid points at which $h(r)$ is evaluated
(with $l$ being a microscopic cutoff length of the order of the width
of the bilayer), and $\vec{q} = (2\pi/L) \left (n_x, n_y \right)$ is
the wavevector with $n_{x,y} = -L/l,\ldots -1,0,1, \ldots L/l$, the
Hamiltonian decouples into a sum of independent harmonic
oscillators. Applying the equipartition theorem to each Fourier mode,
one finds that
\begin{equation}
  \left \langle \abs{h_{\vec q}}^2 \right \rangle = \frac{k_B T
  L^2}{l^4 \left ( \gamma q^2 + \kappa q^4
  \right) },
  \label{eq:helf}
\end{equation}
where $k_B$ is Boltzmann constant and $T$ is the temperatrue. The
accuracy of Eq.~(\ref{eq:helf}) has been demonstrated in numerous
atomistic and coarse-grained (CG) bilayer simulations (see, e.g.,
\cite{Edholm2000,Mark2001,Oded2003,Brown2008}). Because this result is
based on a quadratic approximation of Eq.~(\ref{eq:helf0}), it is
strictly valid only for moderately fluctuating surfaces; i.e., for
$|\vec{\nabla}h(\vec{r})|\ll 1$.  Here we demonstrate that charged
membranes in solutions of pentavalent counterions exhibit strong
thermal height fluctuations, and tend to develop long wavelength
bending instabilities. Membrane in solutions of monovalent counterions
do not exhibit this type of behavior, but rather become mechanically
unstable through a surface tension type mechanism.

\begin{figure}[h]
  \captionsetup{justification=raggedright, singlelinecheck=false}
  \includegraphics[width=0.85\textwidth]{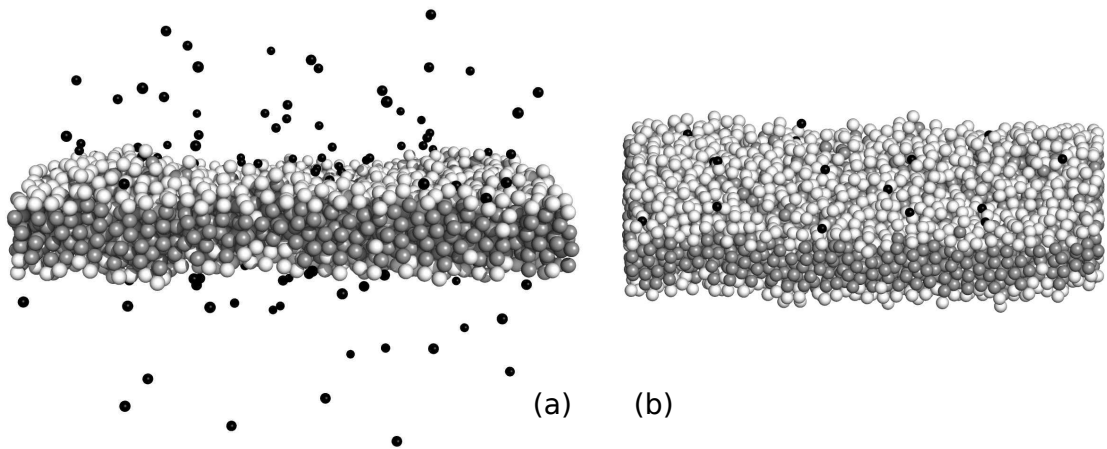}
  \caption{Equilibrium configurations of membranes with charge density
    $\phi=0.08$ in solutions of monovalent (a) and pentavalent (b)
    counterions. The head and tail beads of the lipids appear in
    white and gray colors, respectively, while the ions are presented
    as black spheres.}
  \label{fig:side_dual}
\end{figure}

To allow for simulations of relatively large membranes, we use
Deserno's implicit-solvent model \cite{Deserno2005}, where each lipid
is represented as a trimer consisting of one hydrophilic (head) and
two hydrophobic (tail) beads of size $\sigma$. We set the parameters
of the hydrophobic pair potential to $\epsilon=1.05k_BT$ (potential
depth) and $w_c=1.35\sigma$ (potential range). This choice yields
remarkably soft membranes with bending rigidity $\kappa\simeq 8k_BT$,
which is an essential property for this study since the electrostatic
contribution to $\kappa$ is expected to be small and difficult to
measure ($|\delta \kappa| \simeq k_BT$). We convert a fraction $\phi$
of the neutral lipids to anionic lipids by introducing a charge of
size $-e$ at the center of their head beads. The negative charges of
the lipids are neutralized by either monovalent or pentavalent
counterions, with no added salt. In physical units $\sigma\simeq
6.5{\rm \AA}$, corresponding to a bilayer of thickness
$2\times3\sigma\sim 4$ nm. Setting the temperature to 300K implies
that the Bjerrum length $l_{\rm B}\simeq 7.1{\rm \AA}\simeq 1.1\sigma$. The
short-range repulsive potential between ions and beads is given by the
same form as the bead-bead repulsive potential in Deserno's model (see
Eq.~(1) in Ref.~\cite{Deserno2005}) with $b_{\rm ion-head}=0.5\sigma$
and $b_{\rm ion-tail}=1.5\sigma$.  This choice of parameters allows
the ions to approach the surface of the head beads, while excluding
them from the hydrophobic core of the bilayer. Electrostatic
interactions are computed using Lekner summations \cite{Niels1997}. We
neglect dielectric discontinuities (i.e., image charges) and assume
that all the electrostatic interactions take place in a medium with
uniform dielectric constant $\epsilon=80$ \cite{footnote1}. The
membranes are simulated at zero surface tension using the method
described in Ref.~\cite{Farago2007} for fixed-tension implicit-solvent
bilayer simulations. Initially, we place a flat membrane with
$N/2=1000$ lipids per monolayer in the middle of the simulation box,
with the counterions distributed evenly above and below the membrane.
The system is allowed to equilibrate for $10^5$ Monte Carlo (MC) time
units, where each time unit $\tau$ consists of move attempts to
translate and rotate the lipids, ions displacements, and changes in
the cross-sectional area of the system. To accelerate the relaxation
of the long-wavelength Fourier modes, we also perform a few ``mode
excitation'' moves per $\tau$ \cite{Farago2008}. After initial
thermalization, the systems are simulated for additional
$1.8\times10^6\tau $, and the quantities of interest are sampled every
$50\tau$. Equilibrium configurations of membranes with charge density
$\phi=0.08$ in solutions of monovalent and pentavalent counterions are
displayed in Figs.~\ref{fig:side_dual}(a) and (b), respectively. In
the latter case, the ions tend to condense on the membrane, forming a
thin Gouy-Chapmann ``double layer'' \cite{Safran2002,Netz2002}.

\begin{figure}[h]
  \captionsetup{justification=raggedright, singlelinecheck=false}
  \includegraphics[width=0.825\textwidth]{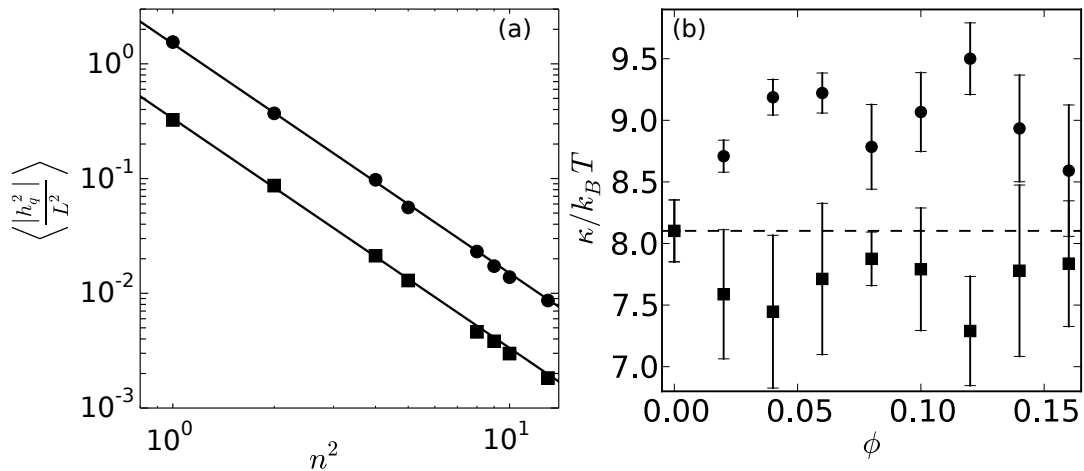}
  \caption{(a) Spectral intensity as a function the squared wavenumber
    $n^2$. Data has been obtained from simulations of charged
    membranes with surface charge density $\phi=0.08$ in solutions of
    monovalent (circles) and pentavalent (squares) counterions. Solid
    line indicates a fit to the power law $\langle|h_q|^2\rangle\sim
    n^{-4}$ based on the four largest Fourier modes. For clarity, we
    vertically shifted the graph corresponding to the monovalent
    counterions simulations, by multiplying the spectral intensities
    by a factor of 5. (b) The bending rigidity $\kappa$ as a function
    of the charge density $\phi$ for membranes with monovalent
    (circles) and pentavalent (squares) counterions. Horizontal dashed
    line indicates the value of $\kappa$ for neutral membranes
    ($\phi=0$).}
  \label{fig:kappa_stable}
\end{figure}

Figure \ref{fig:kappa_stable}(a) depicts the spectral intensity
[Eq.~(\ref{eq:helf})] of the membrane thermal undulations computed for
the bilayers with $\phi=0.08$ whose snapshots are shown in
Fig.~\ref{fig:side_dual}. The graphs have been vertically shifted for
clarity. Both graphs exhibit the power law $\langle|h_q|^2\rangle\sim
n^{-4}$, in agreement with the form of Eq~(\ref{eq:helf}) for
$\gamma=0$. By fitting the simulation results to Eq.~(\ref{eq:helf}),
one can extract the value of $\kappa$.  Our results are summarized in
Fig.~\ref{fig:kappa_stable}(b), showing $\kappa$ as a function of
$\phi$ for membranes in solutions of monovalent (circles) and
pentavalent (squares) counterions. The dashed line denotes the value
of $\kappa$ for a neutral membrane ($\phi=0$). We note that the error
bars on our measurements of $\kappa$ are quite large, reflecting not
only the difficulty in obtaining good statistics for the spectral
intensity of the thermal undulations, but also uncertainties in
fitting the data to the functional form of
Eq.~(\ref{eq:helf}). Therefore, it is impossible to draw quantitative
conclusions from the data regarding the variations of $\kappa$ with
$\phi$. Nevertheless, the data in Fig.~\ref{fig:kappa_stable}(b)
clearly supports the picture that the bending modulus of charged
membranes increases from its value for $\phi=0$ when the counterions
are monovalent. This observation is consistent with the PB theory,
although it should be acknowledged that previous theoretical
calculations of $\delta\kappa$ were done for systems with extra salt
and for stationary (non-undulating) membranes \cite{Mean1995}.

The PB theory is expected to break down when the so called
dimensionless coupling parameter $\Xi=2\pi Z^3l_B^2\phi/a_0$ (where
$a_0$ is the area per lipid and $Z$ denotes the valance of the
counterions) becomes much larger than unity. Given the strong
dependence of $\Xi$ on $Z$, it is not surprising that simulations with
pentavalent counterions reveal a very different trend of reduction in
$\kappa$ due to electrostatic effects. As in the case of monovelent
counterions, the large error bars preclude quantitative analysis of
the variation of $\kappa$ with $\phi$. The observation that the
bending modulus is reduced when the membrane is charged and suspended
in multivalent counterions solution agrees with previous theoretical
studies. The fact that the magnitude of the negative electrostatic
contribution to $\kappa$ is fairly small ($\lesssim 1k_BT$) is also in
general agreement with existing theoretical calculations. As discussed
at the introduction, the negative electrostatic contribution to
$\kappa$ in pentavalent counterions solutions has been attributed to
the attraction due to spatial charge correlations in the double layer,
which allows the membrane to bend more easily~\cite{Sung2002}.

\begin{figure}[b]
  \captionsetup{justification=raggedright, singlelinecheck=false}
  \includegraphics[width=0.5\textwidth]{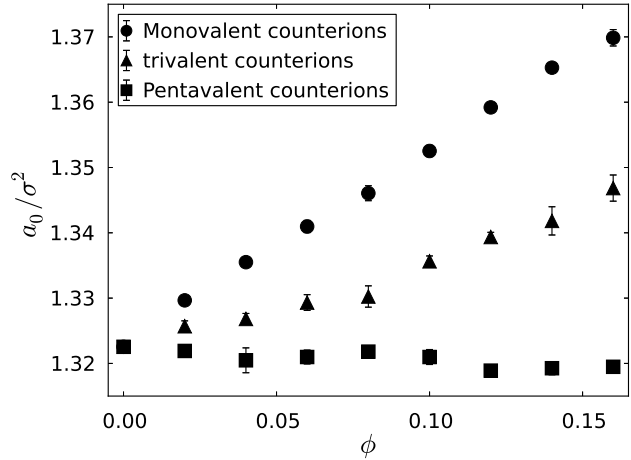}
  \caption{Projected area per lipid $a_0$ as a function of $\phi$ for
    membranes with monovalent (circles), trivalent (triangles), and
    pentavalent (squares) counterions.}
  \label{fig:area}
\end{figure}

The picture emerging from Fig.~\ref{fig:kappa_stable} is also
consistent with our measurements of the equilibrium projected area per
lipid, $a_0=2\langle L\rangle^2/N$, depicted in
Fig.~\ref{fig:area}. (The prefactor 2 in the definition of $a_0$ is
due to the fact that the number of lipids per monolayer is $N/2$.) In
the presence of monovalent counterions, the area per lipid increases
linearly with $\phi$. The increase in $a_0$ arises from the {\em
repulsive} electrostatic interactions between the charged lipids, the
strength of which is enhanced with the increase in the density $\phi$
of the charged lipids. The pentavalent counterion simulations feature
a markedly different behavior, exhibiting a slight decrease in $a_0$
with $\phi$.  The decrease in $a_0$ in this case indicates that the
effective electrostatic interactions between the lipids and
counterions in the double layer become {\em attractive} due to spatial
charge correlations. Also shown in the figure are results of similar
simulations with trivalent counterions which exhibit an intermediate
behavior between the monovalent and pentavalent counterions. The
observed increase in $a_0$ may be attributed to the fact that the
coupling parameter corresponding to the trivalent counterions
simulations satisfies $\Xi\lesssim 10$, which is still within the
range where, usually, mean field theory usually still holds.  The same
trend of ``intermediate'' behavior of trivalent counterions is also
observed in our results for the bending rigidity [data not shown in
Fig.~\ref{fig:kappa_stable}(b)], in which the electrostatic
contribution was found to be vanishingly small.

The increase in area per lipid reported in Fig.~\ref{fig:area} can be
understood as follows: We first note that in the case of no salt, the
PB electrostatic free energy $F_{\rm PB-el}=k_BT\phi N$ (derived by
integrating the electrostatic energy density over the entire space) is
{\em independent}\/ of the surface area $A$. The repulsion arises from
the entropy of the counterions which, to an approximation, can be
viewed as confined within a volume of size $V=Al_{\rm GC}$ around the
surface, where $l_{\rm GC}$ is the Gouy-Chapmann length. The
associated free energy contribution is $F_{\rm PB-en}\sim -k_BT\phi
N\ln(V)=-k_BT\phi N\ln(A^2/N\phi)$, where the last equality is due to
the fact that $l_{\rm GC}\sim(N\phi/A)^{-1}$. Introducing the area per
lipid $a$, and expanding the logarithm around $a^*=a(\phi=0)$, we find
that the area-dependent part of the PB free energy is given by $F_{\rm
PB}=-ck_BT\phi N(a-a^*)/a^*$, where $c$ is a numerical prefactor and
the minus sign accounts for the fact that $F_{\rm PB}$ is repulsive.
Adding $F_{\rm PB}$ to the elastic energy of the uncharged membrane,
yields the following expression for the elastic energy per lipid
$f=F/N$
\begin{equation} 
f=\frac{1}{2}K_A\frac{(a-a^*)^2}{a^*}-ck_BT\phi \frac{(a-a^*)}{a^*},
\label{eq:freeenergy}
\end{equation}
where $K_A$ is the area stretch modulus. This free energy attains a
minimum at the area per lipid: 
\begin{equation}
a_0=a^*+c\phi(k_BT/K_A),
\label{eq:linarea}
\end{equation}
which grows linearly with $\phi$ as depicted in Fig.~\ref{fig:area}.

\begin{figure}[h]
  \captionsetup{justification=raggedright, singlelinecheck=false}
  \includegraphics[width=0.4\textwidth]{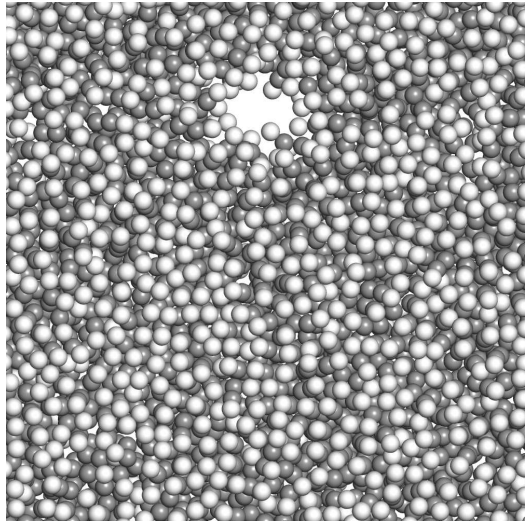}
  \caption{Top view of a simulated membrane with a pore. The membrane,
    with charge density $\phi=0.2$, is in contact with a solution of
    monovalent counterions (not displayed). Color coding is similar to
    Fig.~\ref{fig:side_dual}.}
  \label{fig:hole_mono_400}
\end{figure}

Interestingly, both experiments \cite{petrache} and atomistic
simulations \cite{tieleman} found the area of monovalently charged
phosphatidylserine (PS) lipids to be {\em smaller}\/ than the area of
their neutral phosphattidylcholine (PC) analogs. This counterintuitive
result was primarily attributed to the formation of transient
intra-molecular hydrogen bonds between the amyne and carboxylate
groups of the PS headgroup. Our CG model allows us to ``turn off'' the
hydrogen bonding effect and ``isolate'' the Coulombic contribution,
which turns out to be repulsive in monovalent systems. In our CG
simulations, the repulsive electrostatic interactions are balanced by
relatively soft hydrophobic interactions. Because of the weakness of
these attractive interactions, we observe that for $\phi>0.16$, the
areal strain in the monovalent counterion simulations exceeds the
rupture strain of the bilayer membrane, leading to the formation of
membrane pores, as demonstrated in Fig.~\ref{fig:hole_mono_400}.  The
rupture value of $\phi$ could be increased by including hydrogen
bonding in our CG model or, alternatively, by strengthening the
hydrophobic interactions, but this will also lead to an undesirable
increase in $\kappa$. Real PS bilayers have $\kappa\sim 20-50k_BT$
which is several times larger than of the membrane simulated here.
Assuming a linear relationship between the area stretch modulus $K_A$
and bending rigidity $\kappa$: $\kappa\sim K_Ad^2$, where $d$ is the
bilayer thickness \cite{rawicz}, we can expect the stretch modulus of
real bilayers to also be a few times larger than in simulations. This
feature of real PS bilayers, together with the extra attractive
interaction provided by the hydrogen bonds, explain their mechanical
stability at all charge densities, including for $\phi=1$. The
magnitude of the hydrogen bonding interactions (per lipid) can be
roughly estimated by adopting Eqs.~(\ref{eq:freeenergy}) and
(\ref{eq:linarea}) derived for the case of repulsive electrostatic
interactions, with a modified (negative rather than positive) constant
$c$. For fully charged membrane ($\phi=1$) PS bilayers, the area
stretch modulus is typically $K_A\sim 0.15\ {\rm J/m}^2$, and the
H-bond interactions reduce the area per lipid from $a^*\sim 0.72\ {\rm
nm}^2$ to $a_0\sim 0.65\ {\rm nm}^2$ \cite{petrache}.  Substituting
these values into Eq.~(\ref{eq:linarea}) yields $c\sim -2.5$. Using
this value of $c$ in the second term on the r.h.s.~of
Eq.~(\ref{eq:freeenergy}) gives an estimate for the H-bonding free
energy contribution which is $f_{\rm H-b}\sim-0.25 k_BT$.

\begin{figure}[h]
  \captionsetup{justification=raggedright, singlelinecheck=false}
  \includegraphics[width=0.7\textwidth]{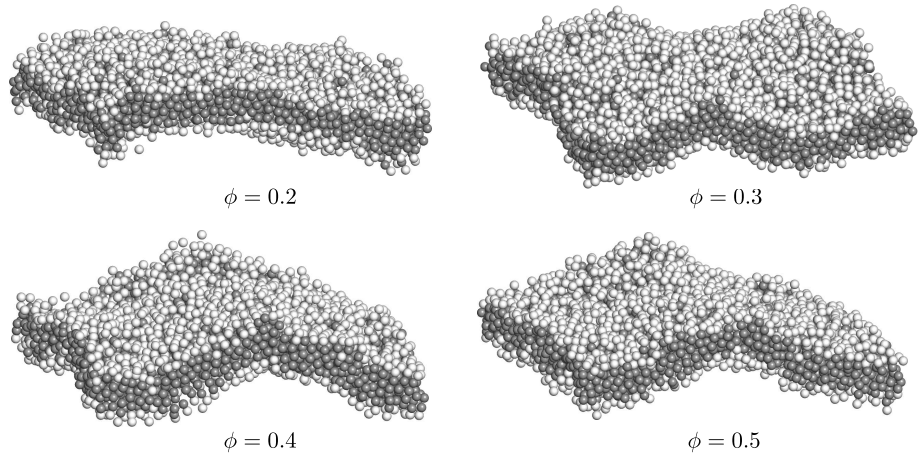}
  \caption{Sample snapshots of strongly undulating charged bilayers
    with pentavalent counterions (not displayed). Color coding is
    similar to Fig.~\ref{fig:side_dual}.}
  \label{fig:bends}
\end{figure}

\begin{figure}[b]
  \sloppy
  \captionsetup{justification=raggedright, singlelinecheck=false}
  \includegraphics[width=0.55\textwidth]{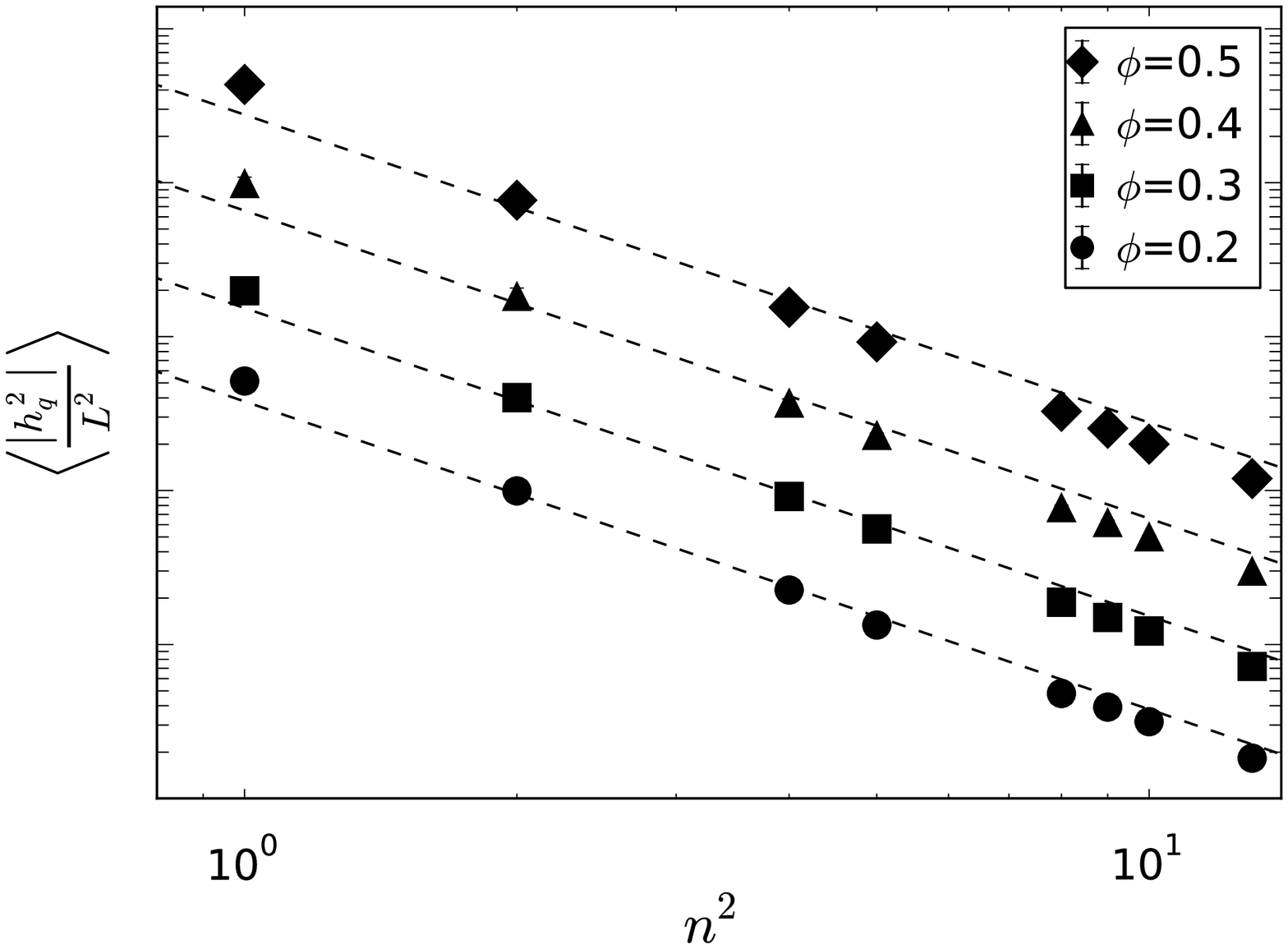}
  \caption{Spectral intensity of membranes in solutions of multivalent
    counterions, with surface charge density $\phi=0.2$ (circles), 0.3
    (squares), 0.4 (triangles), and 0.5 (diamonds). Dashed line
    indicates attempts to fit the data from the second ($n^2 = 2$) and
    third ($n^2 = 4$) largest modes to the power law form
    $\langle|h_q|^2\rangle\sim n^{-4}$. The graphs have been shifted
    vertically for clarity.}
  \label{fig:fit_23}
\end{figure}

Pore formation, as exhibited in Fig.~\ref{fig:hole_mono_400}, is not
observed when the charged membranes are simulated with pentavalent
counterions. Therefore, such membrane can be simulated at much higher
values of $\phi$. However, the multivalent counterion simulations
feature a different type of mechanical instability, which is directly
related to the previously discussed reduction in $\kappa$. At high
charge densities, the membranes in pentavalanet counterion solutions
begin to develop large wavelength bending instabilities, as
illustrated in the series of snapshots in Fig.~\ref{fig:bends},
corresponding to membranes with $0.2\leq\phi\leq 0.5$. The growth in
the amplitude of the undulations, observed in Fig.~\ref{fig:bends},
can also be inferred from the results of Fig.~\ref{fig:fit_23}. Here,
we plot the spectral intensity of the membranes whose snapshots are
displayed in Fig.~\ref{fig:bends}. Clearly, there is poor agreement
between the results in Fig.~\ref{fig:fit_23} and
Eq.~(\ref{eq:helf}). The deviation of the computational results from
Eq.~(\ref{eq:helf}) is expected because the power law
$\langle|h_q|^2\rangle\sim n^{-4}$ is derived from the quadratic
approximation of Eq.~(\ref{eq:helf0}) which, strictly speaking, is
only applicable to weakly fluctuating membranes. The dashed lines
represent attempts to fit Eq.~(\ref{eq:helf}) to the data from the
second ($n^2=2$) and third ($n^2=4$) largest modes. These lines
highlight the rapid increase in the undulation amplitude of largest
Fourier modes ($n^2=1$), which are also the softest modes and the
first to become unstable as $\phi$ increases. The onset of this
bending instability can thus be associated with the decline of the
``apparent bending modulus'' of the first mode, $\kappa_1$, which is
the value of $\kappa$ that solves Eq.~(\ref{eq:helf}) for $n^2=1$. The
results in Fig.~\ref{fig:fit_23} corresponds to $\kappa_1/k_BT=5.1\pm
0.5$, $5.2\pm0.7$, $4.2\pm0.4$, and $3.8\pm0.4$ for $\phi=0.2$, 0.3,
0.4, and 0.5, respectively. We notice that these values of $\kappa_1$
are smaller than the values of $\kappa$ reported in
Fig.~\ref{fig:kappa_stable}(b) for low charge densities. At even
larger charge densities ($\phi>0.5$), we observe that the undulations
continue to grow and ultimately lead to the dissociation of the
bilayer membranes.

To conclude, we investigated the elastic properties of charged
membranes in contact with counterion solutions. The cases of
monovalent and multivalent (with $Z=5$) counterions exhibit distinctly
different behaviors. In the former, both the bending rigidity and the
equilibrium projected area increase with the membrane charge
density. These observations suggest, in agreement with the
Poisson-Boltzmann mean field theory, that the repulsive forces between
the lipid charges are only partially screened by the monovalent
counterions. In the latter case the trends are opposite, namely both
$\kappa$ and $a_0$ show a slight decrease with increasing
$\phi$. These observation can be attributed to the formation of a thin
layer of counterions around the membrane, and the fact that the forces
between spatially correlated charges within the ``double layer''
become attractive. More specifically, the presence of multivalent
counterions creates regions within the double layer where local charge
densities of opposite signs attract each other. The increase in the
curvature undulations and decrease in the area per lipid represent
mechanisms through which the distances between these correlated
regions, especially those residing on the same side of the bilayer,
are generally decreased (see illustration in Fig.~1
ref.~\cite{Shklovskii1999}). The different elastic properties of
membranes in monovalent and multivalent solutions lead to different
mechanical instabilities. In the former case, pores open to relieve
the electrostatic tensile stresses, while the latter case is
characterized by a growth in the amplitudes of large wavelength
bending modes. As a final note we recall that the elastic properties
of real membranes may be affected by other intermolecular forces that
can dominate the electrostatic effect on the bending rigidity. Several
such ``counter mechanisms'' have been mentioned in the text, including
hydrogen-bonding interactions, screening by salt, ions-lipids excluded
volume interactions, and image charges that weaken the binding of the
multivalent counterions to the membrane \cite{buyukdagli}. The CG
simulations provide a framework for systematically exploring the
effects of these additional interactions.

\FloatBarrier 
This work was supported by the Israel Science Foundation
through grant No.~1087/13.


\begin{thebibliography}{99}

\bibitem{Alberts2002} B. Alberts {\em et al.}\/, \emph{Molecular Biology
of the Cell}\/ (Garland Science, New York, 2002).
\bibitem{helf1973} W. Helfrich, Z. Naturforsch. {\bf 28C},
693 (1973).  
\bibitem{Pincus2004} O. Farago and P. Pincus, J. Phys. Chem. {\bf 120},
  2934 (2004).
\bibitem{Mean1995} For a comprehensive review see: D. Andelman in
\emph{Structure and Dynamics of Membranes}\/, Eds. R. Lipowsky and
E. Sackmann (Elsevier, Amsterdam, 1995).
\bibitem{Pincus1998} A. W. C. Lau and P. Pincus, Phys. Rev. Lett. {\bf
81}, 1338 (1998).
\bibitem{Shklovskii1999} T. T. Nguyen, I. Rouzina, and B. I. Shklovskii,
Phys. Rev. E {\bf 60}, 7032 (1999).  
\bibitem{Netz2001} R. R. Netz, Phys. Rev. E {\bf 64}, 051401 (2001).
\bibitem{Sung2002} Y. W. Kim and W. Sung, Europhys. Lett. {\bf 58},
  147 (2002).  
\bibitem{Safran2002} A. W. C. Lau, D. B. Lukatsky, P. Pincus, and
S. A. Safran, Phys. Rev. E {\bf 65}, 051502 (2002).
\bibitem{Netz2002} A. G. Moreira and R. R. Netz, Euro. Phys, J. E {\bf
8}, 33 (2002).
\bibitem{Bloomfield1996} I. Rouzina and V. A. Bloomfield,
J. Phys. Chem. {\bf 100}, 9977 (1996).  
\bibitem{Gelbert1997} N. Gr\o nbech-Jensen, R. J. Mashl,
R. F. Bruinsma, and W. M. Gelbart, Phys. Rev. Lett. {\bf 78}, 2477 (1997).
\bibitem{Beardmore1998} N. Gr\o nbech-Jensen, K. M. Beardmore, and
P. Pincus, Physica A {\bf 261}, 74 (1998).
\bibitem{fleck} C. C. fleck and R. R. Netz,
Phys. Rev. Lett. {\bf 95}, 128101 (2005).

\bibitem{Edholm2000} E. Lindhal and O. Edholm, Biophys. J. {\bf 79},
426 (2000).
\bibitem{Mark2001} S. J. Marrink and A. E. Mark, J. Phys. Chem. {\bf
105}, 6122 (2001).  
\bibitem{Oded2003} O. Farago, J. Chem. Phys. {\bf 119}, 596 (2003).
\bibitem{Brown2008} F. Brown, Annu. Rev. Phys. Chem. {\bf 59}, 685
(2008). 
\bibitem{Deserno2005} I. R. Cooke, K. Kremer, and M. Deserno,
Phys. Rev. E {\bf 72}, 011506, (2005). We slightly modified the model
to avoid occasional escape of lipids from the bilayer.
\bibitem{Niels1997} N. Gr\o nbech-Jensen, G. Hummer, and K. M. Beardmore,
Mol. Phys. {\bf 92}, 941 (1997).
\bibitem{footnote1} It is practically impossible to
calculate image charges for a fluctuating surface (which is the reason
why they are routinely ignored in simulations of many-particle charged
systems). In phospholipid bilayers, the image charges exclude the
electric field from the low dielectric hydrophobic core of the
membrane. However, we note that the exclusion of the electric field
from the membrane is not only a matter of dielectric discontinuities,
but can be also attributed to the nearly flat geometry and the charge
neutrality on both sides of the membrane. Thus, ignoring dielectric
discontinuities may actually be reasonable in some simulations of
interfaces. See more detailed discussion in: R. J. Mashl and N. Gr\o
nbech-Jensen, J. Chem. Phys. {\bf 109}, 4617 (1998).

\bibitem{Farago2007} O. Farago and N. Gr\o nbech-Jensen,
Biophys. J. {\bf 92}, 3228 (2007).
\bibitem{Farago2008} O. Farago, J. Chem. Phys. {\bf 128}, 184105
(2008).

\bibitem{petrache} H. I. Petrach, S. Tristram-Nagle, K. Gawrisch,
D. Harries, V. A. Persegian, and J. F. Nagle, Biophys. J. {\bf 86},
1574 (2004).

\bibitem{tieleman} P. Mukhopadhyay, L. Monticelli, and D. P. Tieleman,
Biophys. J. {\bf 86}, 1601 (2004).

\bibitem{rawicz} W. Rawicz, K. C. Olbrich, T. McIntosh. D. Needham,
and E. Evans, Biophys. J. {\bf 79}, 328 (2000).

\bibitem{buyukdagli} S. Buyukdagli, M. Manghi, and J. Palmeri,
Phys. Rev. E {\bf 81}, 041601 (2010).



\end{thebibliography}
\end{document}